\documentclass[twocolumn,narrowdisplay]{elsart3p}

\usepackage{graphicx,amssymb,citesort}

\journal{Physica B}

\begin{document}
\begin{frontmatter}

\title{Local singlets, frustration, and unconventional superconductivity in the
organic charge-transfer solids\thanksref{grants}}
\thanks[grants]{Supported by DOE Grant No. DE-FG02-06ER46315 and NSF Grant No. DMR-0705163.}

\author[1]{R.T. Clay\thanksref{cor}}
\thanks[cor]{email: r.t.clay@msstate.edu}
\author[2]{S. Mazumdar}
\author[2]{H. Li}

\address[1]{Department of Physics and Astronomy and HPC$^2$ Center for
Computational Sciences, Mississippi State University, Mississippi State MS 39762}
\address[2]{ Department of Physics, University of Arizona
Tucson, AZ 85721}

\date{\today}
\begin{abstract}
We suggest that superconductivity (SC) in the organic charge transfer
solids (CTS) is reached from a Bond-Charge Density Wave (BCDW).  We
discuss an effective model for the BCDW to SC transition, an
attractive $U$ extended Hubbard Hamiltonian with repulsive nearest
neighbor interaction $V$.  We discuss experimental consequences of the
theory for different classes of CTS superconductors as well as related
inorganic materials.
\end{abstract}
\end{frontmatter}

\section{Introduction}

After thirty years of experimental and theoretical effort, the
mechanism of superconductivity (SC) in the organic charge-transfer
solids (CTS) is still elusive.  Electron-electron (e-e) interactions,
electron-phonon (e-p) interactions, and lattice frustration all play
key roles in the unusual insulating phases found in the CTS, but it is
not clear how these apparently very different effects come together in
the SC state. In this paper we outline a unified theoretical approach
to unconventional SC in the entire family of organic CTS, and argue
that all three of the above interactions are essential for SC.  Unlike
existing theories of SC in the CTS, our work can potentially explain the
pairing mechanism in {\it all} $\frac{1}{4}$-filled molecular
superconductors, spanning from the quasi-one-dimensional (1D)
(TMTTF)$_2$X to the nearly isotropic two-dimensional (2D)
$\kappa$-(BEDT-TTF)$_2$Cu$_2$(CN)$_3$ and
EtMe$_3$Z[Pd(dmit)$_2$]$_2$. The theory may additionally explain SC
found in inorganic $\frac{1}{4}$-filled superconductors.

SC in many CTS, for example in $\kappa$-(ET)$_2$X is proximate to
antiferromagnetism (AFM). Further, these 2D CTS are strongly dimerized
suggesting the possibility of describing the material by an effective
$\frac{1}{2}$-filled band (one carrier per dimer).  It has
consequently been proposed that the $\frac{1}{2}$-filled band
repulsive Hubbard model on an anisotropic triangular lattice can
explain the superconducting behavior of the 2D CTS
\cite{Powell05a,Kyung06a}.

The $\frac{1}{2}$-filled band triangular lattice Hubbard model believed to
\begin{figure}[t]
\centerline{\resizebox{2.75in}{!}{\includegraphics{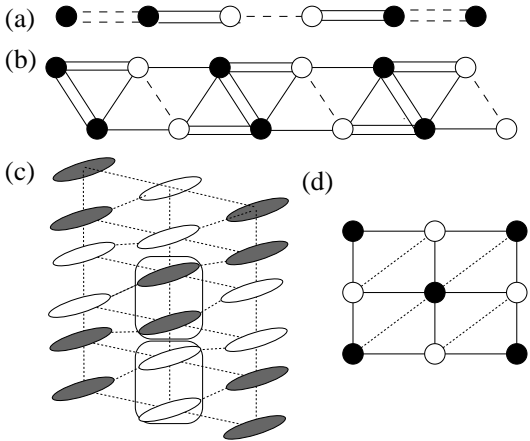}}}
\caption{(a) The ...1100... 1D BCDW state.  Double, double-dashed and
single-dashed lines indicate strong, weak and weakest bonds,
respectively. (b) The zigzag ladder BCDW state. Double, single and
dashed lines again indicate strong, weak and weakest bonds.  (c) 2D
BCDW (Valence-Bond Solid (VBS)) state in the CTS. Filled (open)
ellipses are charge-rich (charge-poor) molecules.  Circled pairs of
sites constitute a single site in the effective negative-$U$ model in
(d), where the filled (open) circles are doubly occupied (vacant)
sites, respectively.}
\label{cartoons}
\end{figure}
show SC has two parameters, the onsite Coulomb interaction, $U$, and
the lattice anisotropy, $t^\prime/t$.  The limit $t^\prime=0$
corresponds to the square-lattice Hubbard model, which is an AFM
insulator for any $U>0$.  With increasing $t^\prime$ the AFM state is
destroyed giving a paramagnetic metal (PM). Mean field and variational
calculations find a narrow d-wave SC region in between AFM and PM
phases \cite{Powell05a,Kyung06a}.

A necessary condition for SC mediated by antiferromagnetic
fluctuations is that $U$ must {\it enhance} the superconducting
pair-pair correlations. We have calculated exact superconducting
pair-pair correlation functions for the $\frac{1}{2}$-filled band
triangular Hubbard model, and find instead that the strength of
pairing correlations always decays monotonically with increasing $U$
\cite{Clay08b}. These results show that there is no SC within this model.

\section{Effective model for BCDW/SC transition}

SC in CTS is proximate to not only AFM but to many other exotic
insulating states such as charge ordering (CO) and spin-Peierls (SP)
states.  We argue that to understand the SC in the CTS, one must first
understand these insulating states. This requires returning to the
full Hamiltonian with the correct band filling ($\frac{1}{4}$-filled
or a carrier density of $n=\frac{1}{2}$).  For the 1D systems our
Hamiltonian contains inter-site e-p
coupling, intra-site e-p coupling, and Hubbard and extended
Hubbard e-e interactions, $H=H_{\rm{inter}}+H_{\rm{intra}}+H_{\rm{ee}}$:
\begin{eqnarray}
\label{hamiltonian}
H_{\rm{inter}}&=& -t\sum_{\langle ij \rangle,\sigma} [1+\alpha(x_{j}-x_{i})](
c^\dagger_{j,\sigma}c_{i,\sigma}+ H.c.) \\
&=&+\frac{1}{2}K_{\rm{inter}}\sum_{\langle ij \rangle} (x_i-x_j)^2 \nonumber \\
H_{\rm{intra}}&=& g\sum_i \nu_i n_i + +\frac{1}{2}K_{\rm{intra}}\sum_{i} \nu_i^2 \nonumber \\
H_{\rm{ee}}&=&U\sum_i n_{i,\uparrow}n_{i,\downarrow} + V \sum_{\langle ij \rangle}
n_in_j \nonumber
\end{eqnarray}
$c^\dagger_{i,\sigma}$ creates an electron of spin $\sigma$ on site
$i$, $x_i$ is the position coordinate of molecule $i$, and
$n_{i,\sigma}=c^\dagger_{i,\sigma} c_{i,\sigma}$,
$n_i=n_{i,\uparrow}+n_{i,\downarrow}$. $\nu_i$ is the coordinate of
the intra-molecular vibration on site $i$, and $K_{inter}$ and $K_{intra}$
are the spring constant for inter-site and intra-site phonons. The density of
carriers is $n=\frac{1}{2}$.

In a series of works, we have investigated this Hamiltonian and its 2D
counterparts
\cite{Clay03a,Clay07a,Mazumdar99a,Mazumdar00a,Clay05a,Clay02a}.  In
general two different kinds of charge ordered states are possible: (i)
Wigner-crystal ..1010.. CO driven by the nearest-neighbor interaction $V$,
and (ii) bond-charge
density-wave (BCDW) ordering with CO pattern ..1100..  driven by the
cooperation of e-e and e-p interactions. In this notation, `1'
indicates a charge density of 0.5+$\delta$, and `0' a charge density
of 0.5-$\delta$. For realistic $U/t \sim 6 - 8$ and $V/t \sim 1 - 3$
the BCDW state tends to dominate
\cite{Clay03a,Clay07a,Mazumdar99a,Mazumdar00a,Clay05a,Clay02a}.
Fig.~\ref{cartoons}(a)-(c) shows the BCDW state for 1D, a zig-zag
ladder lattice, and 2D.

The BCDW in all cases consists of nearest-neighbor singlet
bipolarons.
It gains energy
from the exchange interaction between electrons in the 
pairs and also from e-p interaction via the lattice distortion.
Singlet formation within the BCDW pairs leads to a non-magnetic
ground state in the 1D SP state \cite{Clay07a}, the zigzag
ladder \cite{Clay05a}, as well as the 2D BCDW \cite{Clay02a}.
Such nonmagnetic insulating states coexisting with CO are found
in many CTS.

We map the nearest-neighbor pairs of singly occupied (vacant) sites
into double occupancies (vacancies) as shown in Fig.~\ref{cartoons}(d).
The Hamiltonian describing
this effective model now has a negative  $U$
 and a positive $V$  \cite{Mazumdar08a}:
\begin{figure}
\centerline{\resizebox{3.0in}{!}{\includegraphics{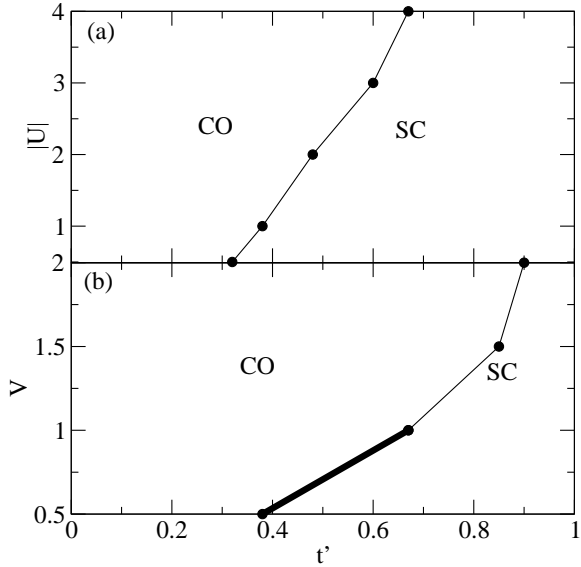}}}
\caption{Phase diagram of the effective model
Eq.~\ref{neguhamiltonian} from exact calculations on a 4$\times$4
cluster. (a) $|U|$-$t^\prime$ phase diagram ($V=V^\prime=1$) (b)
$V$-$t^\prime$ phase diagram ($|U|$=4). The transition is continuous
(thick line) for small $V$.}
\label{negu}
\end{figure}
\begin{eqnarray}
\label{neguhamiltonian}
H_{\rm{eff}} &=& - t\sum_{\langle ij \rangle,\sigma}(c_{i,\sigma}^\dagger c_{j,\sigma}+ H.c.) \\
&-&t^{\prime}\sum_{[kl],\sigma}(c_{k,\sigma}^\dagger c_{l,\sigma}+ H.c.)  \nonumber \\
&-&|U| \sum_{i} n_{i,\uparrow} n_{i,\downarrow} 
+ V\sum_{\langle ij \rangle} n_in_j + V^{\prime}\sum_{[kl]} n_kn_l \nonumber
\end{eqnarray}
Eq.~\ref{neguhamiltonian} describes the effective
lattice of Fig.~\ref{cartoons}(d), with hopping
$t$ along the $x$ and $y$ directions (nearest-neighbor bonds $\langle
ij \rangle$) and $t^\prime$ along the $x+y$ direction. Both $V$ and $V^\prime$
nearest-neighbor Coulomb interactions are included.

Using exact diagonalizations we have shown that the ground state of
Eq.~\ref{neguhamiltonian} is charge ordered for $t^\prime=0$ with the
checkerboard CO of Fig.~\ref{cartoons}(d). As $t^\prime$
increases there is a transition from CO to SC \cite{Mazumdar08a}.
Fig.~\ref{negu} shows the phase diagram as a function of
$t^\prime$-$U$ and $t^\prime$-$V$. The SC pairing in the effective
model is of the onsite singlet form. We note however that in terms of
the original lattice, pairing is primarily nearest-neighbor, and a
full calculation of the original $U>0$
Hamiltonian would likely lead to further pair structure such as nodes
in the gap function.  Furthermore, modifications of the simple
square+$t^\prime$ lattice structure shown in Fig.~\ref{cartoons}(d)
are needed to describe the lattices of specific CTS, for example the
more anisotropic Bechgaard salt superconductors. We believe however
that this effective model can describe the BCDW/SC transition found
across the various CTS families.

\section{Discussion: CTS materials}

\noindent{\it Quasi-1D CTS (TMTSF)$_2$X}: We believe that our theory gives
the correct insight into understanding the SC state in all the
$\frac{1}{4}$-filled CTS superconductors, including both quasi-1D and
2D CTS.  Even for the quasi-1D materials triangular lattice structure is
present in the interchain hopping integrals\cite{Pevelen01a}.
The addition of intrinsic dimerization to Eq.~\ref{hamiltonian}
acts to further stabilize the ..1100.. BCDW over
the ..1010.. CO state\cite{Shibata01a}. 
The role of pressure is to increase interchain coupling and the
effective frustration, giving mobility to the insulating pairs in the
BCDW.  In the quasi-1D (TMTCF)$_2$X (C=Se,S), we have shown that while
the high temperature CO found in this series can have the CO pattern
..1010.., the CO pattern found in the SP state and coexisting with the
SDW adjacent to SC is the ..1100.. BCDW \cite{Clay07a,Mazumdar99a}.

Very large upper critical fields are found in (TMTSF)$_2$X
\cite{Lee97a} and other CTS SC \cite{Zuo00a},
along with an upward curvature of $H_{c2}$ as a function of
temperature. These features are a common characteristic of
superconductivity with short-range local pairing 
and would be expected within our theory\cite{Micnas90a}.
\smallskip

\noindent{\it $\kappa$-(ET)$_2$X:}
While many authors have suggested that the mechanism of
SC in the 2D CTS is linked to the presence of the AFM state,
AFM adjacent to SC is perhaps more unusual than common in the
2D CTS. For example, within the $\kappa$-(ET)$_2$X series,
X=Cu$_2$(CN)$_3$ is a superconductor where the insulating state
is a spin liquid with no apparent spin ordering to millikelvin
temperatures \cite{Shimizu03a,Kurosaki05a}. 
As discussed below, there are many examples of CO/SC transitions
within the CTS; this suggests that the AFM and BCDW states
are quite close in energy, and that under pressure the BCDW
can replace AFM order.

Precision dilatometry measurements on $\kappa$-(ET)$_2$Cu[N(CN)$_2$]Br
find strong lattice effects near the AFM/SC transition \cite{Souza07a}. 
 Magneto-optical measurements on $\kappa$-(ET)$_2$Cu(SCN)$_2$
find changes in the vibrational spectra between SC and normal states,
suggesting that intramolecular vibrations are involved in the
superconducting transition \cite{Olejniczak03a}.
These experiments suggest strongly the importance of lattice
degrees of freedom in the SC transition even when AFM is
present, in agreement with our theory.
\smallskip

\noindent{\it Dmit salts:}
 EtMe$_3$Z[Pd(dmit)$_2$]$_2$ has a charge/bond
ordered insulating state that has been described as valence-bond solid
(VBS) ordering \cite{Tamura06a,Shimizu07a}. We point out that the charge
ordering found in the VBS is the same as in our  ..1100.. BCDW,
see Fig.~3 in reference \cite{Tamura06a}.
Under pressure, the VBS state becomes superconducting.
\smallskip

\noindent{\it $\beta$-(meso-DMBEDT-TTF)$_2$PF$_6$:} 
$\beta$-(meso-DMBEDT-TTF)$_2$PF$_6$ is another 2D CTS that has CO
and undergoes a CO/SC transition under pressure \cite{Kimura06a}.
The CO state has been described as a ``checkerboard'' pattern,
implying that the local CO is ..1010.. driven by the $V$ interaction.
We point out however, that the checkerboard here is in terms
of {\it dimers}, and the CO pattern again corresponds to the BCDW
shown in Fig.~\ref{cartoons}(c).

\section{Relationship to other theories}

Our theory has strong similarities with RVB and bipolaron models of SC
simultaneously.  As in the RVB model, pairing is (partly) driven by
AFM correlations also in our model.  The key difference in our model
is the filling: at density $n=\frac{1}{2}$ because of the fewer number
of neighbors it is much favorable energetically to form a
nearest-neighbor singlet state.  Furthermore, at $n=\frac{1}{2}$ e-e
and e-p interactions tend to act {\it cooperatively}
\cite{Clay03a,Clay07a,Mazumdar99a,Mazumdar00a,Clay05a,Clay02a}.

Traditional bipolaronic SC is based on over-screening of the e-e
interaction by strong e-p interactions \cite{Micnas90a}. This leads to
an unphysically large effective mass of the paired electrons.  In
contrast, because of the {\it co-operation} between the AFM and e-p
interactions, no overscreening is required in our theory.
Furthermore, it has recently been shown that bipolarons are especially
mobile on triangular lattices \cite{Hague07a}.  Both $n=\frac{1}{2}$
and frustration are hence essential ingredients of the
insulator-to-superconductor transition in the CTS.

\section{Application to other materials}

We believe that our theory may be relevant to other $\frac{1}{4}$-filled
superconductors. Several of these materials were known
before superconducting CTS. LiTi$_2$O$_4$ with $\frac{1}{4}$-filled Ti
bands is one of the first oxide superconductors discovered with at
$T_c$ of 12 K \cite{Johnston76a}. Because of the presence of
frustration in the spinel lattice structure, a resonating valence bond
(RVB) ground state has been suggested \cite{Satpathy87a}.  However,
e-p coupling also appears to be strong in this material, and it is
still unclear of the relative importance of e-e and e-p interactions
\cite{Sun04a}.

Very complicated CO ordering occurs in other 
$\frac{1}{4}$-filled spinels and is in fact a combination of orbital
and charge ordering. In CuIr$_2$S$_4$, CO and a nonmagnetic
(spin singlet) ground state are found. This state has been 
explained as an orbitally driven Peierls state, with the
pattern Ir$^{3+}$-Ir$^{3+}$-Ir$^{4+}$-Ir$^{4+}$ \cite{Khomskii05a}.
This is again the ..1100.. BCDW CO pattern.

A CO/SC transition was found under pressure in
$\beta$-Na$_{0.33}$V$_2$O$_5$ \cite{Yamauchi02a}.  While the
stoichiometry of $\beta$-Na$_{0.33}$V$_2$O$_5$ gives $\frac{1}{12}$
filling of vanadium sites on average, there occur here three
inequivalent vanadium sites with unequal electron
distribution. Experiment indicates $\frac{1}{4}$-filled zig-zag chains
as in the quasi-1D CTS \cite{Okazaki04a}.  The SC state in this
material is extremely sensitive to the concentration of Na, with the
SC transition disappearing for a tiny change in Na concentration
($x$=0.32 rather than $x$=0.33) \cite{Yamauchi08a}.  This strongly
suggests that the CO/SC transition here is a property of the exact
band filling of the system, rather than being driven by doping where a
broad peak around the optimal $x$ would be expected.

\end{document}